\documentclass{jfm}

\usepackage{graphicx}
\usepackage{newtxtext}
\usepackage{newtxmath}
\usepackage{natbib}
\usepackage{subfigure}
\usepackage{hyperref}
\hypersetup{
    colorlinks = true,
    urlcolor   = blue,
    citecolor  = black,
}

\newcommand{\RomanNumeralCaps}[1]

\title{A local intermittency based Reynolds-averaged transition model  for turbulent mixing induced by interfacial instabilities}

\author{Hansong Xie\aff{1},
  Mengjuan Xiao\aff{2}, Yousheng Zhang\aff{1,2}\corresp{\email{zhang\_yousheng@iapcm.ac.cn}}
 \and Yaomin Zhao\aff{1,3}\corresp{\email{yaomin.zhao@pku.edu.cn}}}

\affiliation{\aff{1}HEDPS, Center for Applied Physics and Technology, and College of Engineering, Peking University, Beijing 100871, China
\aff{2}Institute of Applied Physics and Computational Mathematics, Beijing 100094, China
\aff{3}State Key Laboratory for Turbulence and Complex Systems, College of Engineering, Peking University, Beijing 100871, China}

\begin{document}
\maketitle

\begin{abstract}
Accurate prediction of mixing transition induced by interfacial instabilities is vital for engineering applications, but has remained a great challenge for decades. For engineering practices, Reynolds-averaged Navier-Stokes simulation (RANS) is the most viable method. However, existing RANS models for mixing problems are mostly designed for fully developed turbulence, failing to depict the locally spatio-temporal-dependent characteristic of transition. In the present study, the idea of the intermittent factor (denoted as $\gamma$), which has been widely used in boundary layer transition in aerospace engineering, is extended to the mixing problems. Specifically, a transport equation for $\gamma$ is built based on local flow variables, which is used to describe the locally spatio-temporal-dependent characteristic of transition. Furthermore, $\gamma$ is coupled into the widely used K-L turbulent mixing model to constrain the two key product sources terms that dominate the evolution of mixing, i.e. the Reynolds stress and the buoyancy effect. Subsequently, the simulations of two reshocked Richtmyer-Meshkov mixing cases with remarkable transition effects confirm that the proposed model has a good performance for predicting mixing transition. To the best of our knowledge, it is the first study that an extra transport equation for intermittent factor has been proposed for a RANS mixing transition model. More importantly, the present modeling framework is flexible and has the potential to be applied to other RANS models. It provides a promising strategy for more advanced modeling for mixing transition.  
\end{abstract}



\section{Introduction}
\label{sec:Introduction}
Rayleigh-Taylor, 
Richtmyer-Meshkov 
and Kelvin-Helmholtz 
instabilities are crucial in a wide range of natural phenomena and engineering applications, such as supernova explosion \citep{burrows2000supernova}, inertial confinement fusion \citep{thomas2012drive}, and so on. 
In practical flows, these instabilities frequently interact with each other and cause mixing between different materials. 
Therefore, accurate prediction of mixing is of great significance for understanding natural phenomena and optimizing engineering applications.
Three primary methodologies are utilized to numerically predict the evolution of mixing: direct numerical simulation, large eddy simulation (LES), and Reynolds-averaged Navier-Stokes simulation (RANS). Specifically, RANS is widely applied in engineering owing to its computational efficiency.

The mixing evolution generally goes through three stages: development of instability, mixing transition and turbulent mixing.
In the first stage, hydrodynamic instabilities amplify perturbations at the interface, leading to a continuous increment of the interface area.
The instability development during this stage is characterized as linear or weakly nonlinear, owing to the small perturbation amplitudes.
Thus accurate prediction can be achieved through analytical theories or numerical simulations \citep{Liu2022A,zhang2022quan,Mikaelian1998Ana,Layzer1955On}.
Perturbations dominate the flow in the first stage, until the amplitude of interface perturbations becomes comparable to the dominant wavelength.
As the flow transitions to the second stage, the interfacial structures start to break into smaller scales, and multiscale behaviour appears due to the coupling and competition between different modes. 
Subsequently, the flow in the third stage evolves into a well-mixed state and displays self-similarity and deterministic statistical laws, making it suitable for  RANS modeling \citep{zhang2020Methodology,Dimonte2006K-L,Kokkinakis2015Two}. 
Unlike the first and third stages, which can be reasonably well predicted utilizing current methodologies, the profoundly nonlinear dynamics encountered in the intermediate mixing transition stage pose formidable challenges for modeling endeavors.
Specifically, the mixing transition is a temporal evolutionary process characterized by pronounced spatial variations in  different locations. 
This locally spatio-temporally dependent characteristic cannot be accurately captured by existing RANS mixing models, which are typically proposed for fully developed turbulent flows.

For accurate prediction of transition, two key aspects need to be considered. 
Firstly, it is essential to accurately identify the onset of transition. 
Based on observations from a series of flows, \citet{dimotakis2000the} proposed a transition threshold based on the outer-scale Reynolds number, approximately at $Re=1\sim2\times10^4$.
This threshold for transition is established based on the appearance of an inertial range, marking an abrupt transition from a low-mixed state to a well-mixed state. It demonstrates that this threshold actually corresponds to the ultimate Reynolds number attained at the end of the transition. \citet{zhou2003onset} extended Dimotakis's criterion to the mixing problem by introducing an additional length scale that evolves with time.
Notably, the aforementioned criteria, depending on the identification of inertial range, can only be applied \emph{a posteriori}, failing to meet the requirement of real-time predictions. 
Moreover, the corresponding transition threshold value relies on the outer Reynolds number, which is essentially a global quantity unable to capture the local characteristics of the transition.
Recently, \citet{wang2022the} proposed a new transition criterion based on mixing mass, enabling identification of transition onset and cessation times.
Nonetheless, the mixing mass remains a global parameter, rendering it unsuitable for real-time RANS predictions.


Secondly, accurate prediction of the subsequent flow development after the onset of transition is crucial. 
Given that most of the existing RANS mixing models are constructed under the assumption of fully developed turbulence, utilizing these models directly to predict transition would result in an overestimation of the mixing evolution.
To employ RANS models for predicting transitions, \citet{Haines2013Rey} artificially controlled the moment of switching on the turbulence model, based on the timing of shock waves impacting on the interface.
While this approach successfully predicts the evolution of mixing transition, it has an empirical nature. 
A more effective strategy would be to enable the model to automatically characterize the onset and evolution of transition.
To achieve a precise depiction of the transition process, \citet{Grinstein2020Dynamic} proposed a dynamic hybrid RANS/LES bridging approach, aiming to utilize LES information to improve RANS predictions. 
Despite the method's promising potential, it remains nascent and needs further refinement prior to its suitability for engineering applications.

As previously mentioned, the actual transition process exhibits significant local spatio-temporal dependence, necessitating that RANS models possess the capability to capture this characteristic.
It is well known that the intermittent factor $\gamma$, widely employed for boundary layer transition phenomena in aerospace engineering, can reflect this characteristic and is widely used to describe variations of flow states.
Nevertheless, no research has been reported applying this well-developed idea to mixing problems.
Therefore, in this study we extend the concept of the intermittent factor to mixing flows and develop a RANS mixing transition model. As demonstrated later, the current strategy proves to be effective.

The paper is organized as follows. Section \ref{Sec:Methodology} provides a detailed documentation of the modeling process, encompassing the governing equations and the baseline model in Section \ref{Sec:Governing equations and baseline model}, the key modeling idea in Section \ref{Sec:Modelling strategy for the mixing transition model}, and the construction of the transport equation for 
$\gamma$ in Section \ref{Sec:Transport equation for the intermittent factor}. Section \ref{Sec:Model validation} validates the proposed model using two representative cases. Finally, Section \ref{Sec:Conclusion and discussion} summarizes the current study and presents further discussions.

\section{Methodology }\label{Sec:Methodology}

\subsection{Governing equations and baseline model}\label{Sec:Governing equations and baseline model}
The multicomponent RANS equations are solved considering  molecular
transport and thermodynamic coefficients. The transport equations for the mean density $\rho$, velocity $u_{i}$, total energy $E$ of the mixture, and mass fraction $Y_{\alpha}$  of specie $\alpha$ are given as follows:
\begin{eqnarray}
&&\frac{\partial \bar{\rho }}{\partial t}+\frac{\partial \bar{\rho }{{{\tilde{u}}}_{j}}}{\partial {{x}_{j}}}=0,  \label{eq1} \\ 
&&\frac{\partial \bar{\rho }{{{\tilde{u}}}_{i}}}{\partial t}+\frac{\partial \bar{\rho }{{{\tilde{u}}}_{i}}{{{\tilde{u}}}_{j}}}{\partial {{x}_{j}}}+\frac{\partial \bar{p}}{\partial {{x}_{i}}}=-\frac{\partial \tau_{ij}}{\partial {{x}_{j}}}+   
\frac{\partial {{{\bar{\sigma}}}_{ij}}}{\partial {{x}_{j}}}, \label{eq2} \\
&&\frac{\partial \bar{\rho }\tilde{E}}{\partial t}+\frac{\partial (\bar{\rho }\tilde{E}+\bar{p}){{{\tilde{u}}}_{j}}}{\partial {{x}_{j}}}=D_{E}+D_{K}+\frac{\partial }{\partial {{x}_{j}}}\left(-{\tau}_{ij}{{\tilde{u}}_{i}}+{{\bar{\sigma }}_{ij}}{{\tilde{u}}_{i}}   
 -\bar{q}_{c}-\bar{q}_{d}
\right),  \label{eq3} \\
&&\frac{\partial \bar{\rho }{{{\tilde{Y}_{\alpha}}}}}{\partial t}+\frac{\partial \bar{\rho }{{{\tilde{Y}_{\alpha}}}}{{{\tilde{u}}}_{j}}}{\partial {{x}_{j}}}=\frac{\partial }{\partial {{x}_{j}}}\left(-\bar{\rho}\widetilde{u_{i}^{''}{Y_{\alpha}}^{''}}+\bar{\rho}\bar{D}\frac{\partial\tilde{Y}_{\alpha}}{\partial x_{j}}\right). \label{eq4}
\end{eqnarray}
The overbar and tilde symbols represent the Reynolds and Favre averaged fields, respectively.
The double prime symbol denotes Favre fluctuations.
The heat flux $\bar{q}_{c}$ is defined by the Fourier's law as $\bar{q}_{c}=-\bar{\kappa} \partial \tilde{T}/\partial x_{j}$. The interspecies diffusional heat flux $\bar{q}_{d}$ is given by 
$\bar{q}_{d}=-\sum \bar{\rho} \bar{D}C_{p\alpha}\tilde{T}\partial \tilde{Y}_{\alpha}/\partial x_{j}$.
The  viscous stress tensor is given as 
\begin{equation}
\bar{\sigma}_{ij}= 2\bar{\mu}(\tilde{S}_{ij}-\tilde{S}_{kk}\delta_{ij}/3), \quad  \tilde{S}_{ij}=(\partial \tilde{u}_{i}/\partial x_{j}+\partial \tilde{u}_{j}/\partial x_{i})/2. 
\end{equation}
Here $\bar{\mu}$, $\bar{D}$,  $\bar{\kappa}$, and $C_{p\alpha}$ represent dynamic viscosity, mass diffusivity, thermal conductivity, and constant-pressure specific heat of specie $\alpha$, respectively. 
Moreover, $\tau_{ij}$ is the Reynolds stress, and the terms $D_{E}=-\partial(\bar{\rho}\widetilde{{u}_{j}^{''}e^{''}}+\overline{pu_{j}^{''}})/\partial x_{j}$,  $D_{K}=-\partial(\bar{\rho}\widetilde{{u}_{i}^{''}{u}_{i}^{''}{u}_{j}^{''}}/2)/\partial x_{j}$,  and $-\bar{\rho}\widetilde{u_{i}^{''}{Y_{\alpha}}^{''}}$ represent the turbulent diffusion terms of the total energy, turbulent kinetic energy (TKE) $\tilde{K}$, and mass fraction, respectively.  
It should be emphasized that Eqs.(\ref{eq1})$\sim$(\ref{eq4})  are deduced based  on the concept of ensemble averaging and are theoretically applicable to the three stages of mixing evolution.
Once the unclosed terms are appropriately modeled, the equation array can be solved by coupling it with the equation of state (EOS) $\bar{p}M=\bar{\rho}R\tilde{T}$ for the perfect gas, where $M$ and $R$ denoting the molar mass and the gas constant, respectively.
The EOS for the mixture is calculated under the assumptions of iso-temperature and partial-pressure. 
Additionally, the fluid properties of the mixture are determined using a species-linearly weighted assumption \citep{Livescu2013Num}.

In previous studies, the majority of RANS simulations focus on fully developed turbulence. 
Consequently, numerous turbulent mixing models have been proposed, including the K-$\epsilon$, K-L, K-L-a, and Besnard-Harlow-Rauenzahn (BHR) models. In this study we take the well-developed K-L model as an example to briefly illustrate how turbulent flow is modeled. Specifically, the turbulent transport terms are modeled by the gradient diffusion assumption (GDA)
\begin{equation} \label{eq6}
-\bar{\rho}\widetilde{u_{i}^{''}f^{''}}=\frac{\mu_{t}}{N_{f}}\frac{\partial \tilde{f}}{\partial x_{i}}, 
\end{equation}
where  $f$ denotes an arbitrary physical variable and $N_{f}$ is a model coefficient. The symbol $\mu_{t}$ is the turbulent viscosity, which is described by the TKE $\tilde{K}$ and the turbulent length scale $\tilde{L}$ 
\begin{equation} \label{eq7}
\mu_{t}=C_{\mu}^{R}\bar{\rho} \tilde{L}\sqrt{2\tilde{K}}.   
\end{equation}
Here $C_{\mu}^{R}$ is a model coefficient, which is calculated based on the realizability principle \citep{Xiao2021realizability}.
With the Boussinesq eddy viscosity hypothesis, the  Reynolds stress is modeled as 
\begin{equation} \label{eq8}
  {{{\tau }}_{ij}}={{C}_{P}}\bar{\rho }\tilde{K}{{\delta }_{ij}}-2{{\mu }_{t}}(\tilde{S}_{ij}-\tilde{S}_{kk}{{\delta }_{ij}}/3),
\end{equation}
where $C_{P}$ is a  model coefficient. Additionally, the closed transport equations of the TKE and $\tilde{L}$  are 
\begin{eqnarray}
\label{eq9}
&&\frac{\partial \bar{\rho }\tilde{K}}{\partial t}+\frac{\partial \bar{\rho }{{{\tilde{u}}}_{j}}\tilde{K}}{\partial {{x}_{j}}}=-{{{\tau }}_{ij}}\frac{\partial {{{\tilde{u}}}_{i}}}{\partial {{x}_{j}}}+\frac{\partial }{\partial {{x}_{j}}}(\frac{{{\mu }_{t}}}{{{N}_{K}}}\frac{\partial \tilde{K}}{\partial {{x}_{j}}})+{{S}_{Kf}}-{{C}_{D}}\bar{\rho }{{\left( \sqrt{2\tilde{K}}\right)}^{3}}/{\tilde{L}},  \\
\label{eq10}
&&\frac{\partial \bar{\rho }\tilde{L}}{\partial t}+\frac{\partial \bar{\rho }{{{\tilde{u}}}_{j}}\tilde{L}}{\partial {{x}_{j}}}=\frac{\partial }{\partial {{x}_{j}}}(\frac{{{\mu }_{t}}}{{{N}_{L}}}\frac{\partial \tilde{L}}{\partial {{x}_{j}}})+{{C}_{L}}\bar{\rho }\sqrt{2\tilde{K}}+{{C}_{C}}\bar{\rho }\tilde{L}\frac{\partial {{{\tilde{u}}}_{j}}}{\partial {{x}_{j}}},
\end{eqnarray}
where ${{S}_{Kf}}$ is the buoyancy product term.
More details about the K-L model, including the value of the model coefficients and the expression of $S_{Kf}$, can be found in \citet{zhang2020Methodology}. 
Nevertheless, the aforementioned K-L model is not suitable for the mixing transition stage since all closures are developed for fully developed turbulence.
The next two sections will introduce how to construct the mixing transition model on the basis of the baseline K-L model.

\subsection{Modeling strategy for  the mixing transition model} \label{Sec:Modelling strategy for the mixing transition model}
In the transition process, the flow state undergoes dramatic variations and exhibits significant local spatio-temporal dependence. Hence, it is necessary to seek for a suitable flow variable to capture this characteristic.
Exactly, this aligns with the concept of the intermittent factor  $\gamma$, which has been extensively employed to predict boundary layer transitions in aerospace engineering.
Therefore, in this study we extend the idea of the intermittent factor to mixing problems. 

It is important to note that the  $\gamma$ in the boundary layer is often defined based on the time-averaged operations because of the statistically stationary nature, which cannot be directly applied to the unsteady mixing flows. 
Consequently, following the original definition \citep{dopazo1977on}, the concept of the intermittent factor is defined using the ensemble-averaged approach.
Specifically, we define the intermittent function $I(x,y,z,t,n)$, which takes into account time and space variables, as well as the sample parameter $n$. 
Thus, for a given sample, the flow state at a  specified spatio-temporal point is either laminar or turbulent. 
We further assign $I\!=\!0$ for laminar flow and $I\!=\!1$ for turbulence. With the ensemble-averaged approach, the intermittent factor is defined  as 
\begin{equation} \label{eq11}
    \gamma\equiv\frac{1}{N}\sum_{n=1}^{N}I(x,y,z,t,n)=\gamma(x,y,z,t), 
\end{equation}
where $N$  represents the sample size. 

The Reynolds stress plays a crucial role in the transition to turbulence. However, the existing closure is only applicable to turbulent flows, thus necessitating remodeling. In accordance with the definition (\ref{eq11}), we express the new Reynolds stress $\tau_{ij}^{new}$  as 
\begin{equation} \label{eq12}
\tau_{ij}^{new}=(1-\gamma)\tau_{ij}^{lam}+\gamma\tau_{ij}^{tur}.  
\end{equation}
Here $\tau_{ij}^{lam}$ and $\tau_{ij}^{tur}$ represent the Reynolds stresses associated with the laminar and turbulent states, respectively, with $\tau_{ij}^{lam}$ being negligible. 
Thus the new Reynolds stress is only related to the component of the turbulent state $\tau_{ij}^{tur}$, which is closed by Eq.(\ref{eq8}).  
Given that the normal stress component of the Reynolds stress can be determined by TKE, $\gamma$ is only applied to the turbulent viscosity $\mu_{t}$ to modify the deviatoric stress part, i.e.
\begin{equation} \label{eq13}
\tau_{ij}^{new}={{C}_{P}}\bar{\rho }\tilde{K}{{\delta }_{ij}}-2{{\mu }_{new}}(\tilde{S}_{ij}-\tilde{S}_{kk}{{\delta }_{ij}}/3), \quad \mu_{new}=\gamma\mu_{t}=C_{\mu}^{R}\gamma\bar{\rho}\tilde{L}\sqrt{2\tilde{K}}.
\end{equation}
Accordingly, in the turbulent diffusion term modeled by the GDA, $\mu_{t}$  is replaced by $\mu_{new}$.

In addition to the Reynolds stress, the buoyancy product term $S_{Kf}$ in Eq.(\ref{eq9}) is also crucial for the mixing evolution, which also requires further improvement for the mixing transition stage.
To modify the baseline model minimally, the simplest approach is multiplying $S_{Kf}$ by $\gamma$. Nevertheless, this treatment may cause that the TKE cannot be excitated especially for the initial stage of instability due to the quite small value of the intermittent factor $\gamma$. 
Therefore, it is suggested to introduce a small excitation source denoted as  $\zeta\!=\!1\times10^{-9}$ into the buoyancy product term, which dominates the production of TKE. Thus the new buoyancy product term  $S_{Kf}^{new}$ is expressed as   
\begin{equation} \label{14}
S_{Kf}^{new}=max(\gamma,\zeta) S_{Kf}.    
\end{equation}

\subsection{Transport equation for the intermittent factor}\label{Sec:Transport equation for the intermittent factor}
Sec.\ref{Sec:Modelling strategy for the mixing transition model} presents the two key improvements in constructing the mixing transition model. In this section, the transport equation for $\gamma$ is built to describe the local spatio-temporal dependence of the the mixing transition flow and  improve the generalization performance of the models.

Following the framework \citep{wang2011deve} of constructing the boundary layer transition models based on the intermittent factor, we present the transport equation for $\gamma$:
\begin{equation} \label{eq15}
\frac{\partial \bar{\rho }\gamma}{\partial t}+\frac{\partial \bar{\rho }{{{\tilde{u}}}_{j}}\gamma}{\partial {{x}_{j}}}=\frac{\partial }{\partial {{x}_{j}}}\left[\left(\bar{\mu}+\frac{{{\mu }_{eff}}}{{{N}_{\gamma}}}\right)\frac{\partial \gamma}{\partial {{x}_{j}}}\right]+P_{\gamma}-\epsilon_{\gamma}. 
\end{equation}
Here $P_{\gamma}$ and $\epsilon_{\gamma}$ represent the product and dissipation terms, respectively. The model coefficient $N_{\gamma}$ is set to 1. The physical process of  mixing evolution tells that the flow eventually reaches a fully-developed turbulent state, indicating that the Reynolds stress $\tau_{ij}^{new}$ and the buoyancy product term $S_{Kf}^{new}$ should ultimately approach the closed form in the baseline K-L turbulent model. 
Correspondingly, $\gamma$ should monotonically increase to 1 until the flow becomes fully developed turbulence.  This implies that $P_{\gamma}-\epsilon_{\gamma}$ should be greater than 0 to provide a positive net increment. To meet this requirement, we adopt the simplest approach  of setting $\epsilon_{\gamma}\!=\!\gamma P_{\gamma}$.

A transition model must have two basic functions: identifying the onset of transition and predicting the subsequent flow evolution accurately. In the current model, these functions are represented as:
\begin{equation} \label{eq16}
P_{\gamma}=F_{onset}G. 
\end{equation}
Here $F_{onset}$ serves as a transition switch, while $G$ describes the growth rate of $\gamma$. The expression for $F_{onset}$ is given by    
\begin{equation}\label{eq17}
    F_{onset}=1-exp\left(-\frac{C\tilde{L}\sqrt{2\tilde{K}}/\nu}{Re_{tra}}\sqrt\frac{{\left|\nabla E_{u}\right|}}{{\left|\nabla \tilde{K}\right|}}\right). 
\end{equation}
In this equation, $\nu$ and $E_{u}$ represent the molecular kinematic viscosity and the mean kinetic energy respectively, while $C$ is a model coefficient and $Re_{tra}$ corresponds to the ultimate Reynolds number attained at the end of the transition stage, taken as $10^4$ \citep{dimotakis2000the}.
Eq.(\ref{eq17}) is easy to understand with the following explanation. 
Through an analysis of the order, the exponent part in Eq.(\ref{eq17}) can be expressed as $C\tilde{L}\tilde{u}/\nu Re_{tra}$, which is further simplified as $Re_{local}/ Re_{tra}$. 
It is evident that when the flow remains stagnant or the turbulence is weak, $Re_{local}$ approaches $0$, resulting in $F_{onset}\rightarrow0$, which means that transition does not occur. 
Consequently, the intermittent factor does not grow. 
As instabilities develop and fluctuations intensify, mixing begins to evolve.
Once the ratio between the local Reynolds number $Re_{local}$ and the ultimate Reynolds number $Re_{tra}$ surpasses a certain threshold value, the model identifies the occurrence of transition.  
Although there may not be an universal value for the critical ratio across different problems, $0.01$ is a relatively reasonable order. Based on the analysis of the order, the coefficient $C$ should be of the order $1$, and specifically, it is set to $5$ in the present study.



Furthermore, the growth rate of the intermittent factor $\gamma$ is described using the function $G$. 
Since there are no relevant studies available as a reference, we adopt the growth rate of the intermittent factor in boundary layer as a basis.
Accordingly, $G$ is expressed as: 
\begin{equation} \label{eq18}
G=\bar{\rho}S{\gamma}^{0.5}(1-\gamma). 
\end{equation}
Here $S$ represents the amplitude of the mean strain rate, defined as $S\!=\!\sqrt{2\tilde{S}_{ij}\tilde{S}_{ij}}$.





\section{Model validation}\label{Sec:Model validation}
\begin{figure}
    \centering \includegraphics[width=0.9\textwidth,height=0.28\textwidth,]{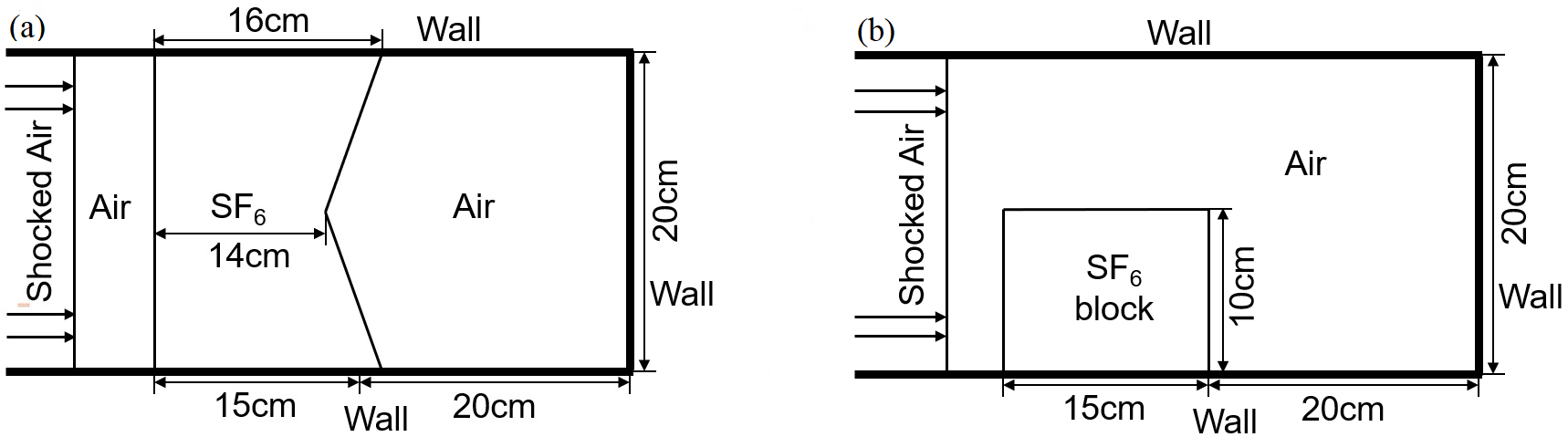}
    \caption{Schematic diagrams of the two reshocked RM mixing cases.}
    \label{fig1}
\end{figure}

In this section, the performance of the proposed model is verified by two representative reshocked RM mixing cases, which originate from the experiments conducted at the linear shock tube facility of AWE (Aldermaston). 
Fig.\ref{fig1}(a) represents the inverse chevron case (noted as case A) in \citet{Hahn2011Richtmyer}, which features a dense SF6 gas block encased in air within a tube. 
Moreover, Fig.\ref{fig1}(b) represents the shock tube experiments (noted as case B) reported in \citet{Bates2007rich}, in which shock waves interact with a half-height rectangular block of dense SF6 gas. 
In both cases, apart from the RM effect, the special interface design leads to an initial KH instability that promotes the transition to turbulence.
Additionally, due to the presence of reflective boundary, the reflected shock wave from the right wall will reshock the mixing region to enhance the mixing and  make the transition effect more conspicuous. 

Because both of the cases have a good two-dimensional (2-D) statistical characteristic, the present RANS calculations are set as 2-D.
The numerical scheme employed is the same as that used in \citet{Xiao2022exp}, with the exception of setting the Courant-Friedrichs-Lewy (CFL) number to 0.05 to enhance computational robustness.
Uniform grids are adopted for both cases, with grid scales $\triangle x$  set to $1/16cm$ and $1/8cm$ for case A and B, respectively. 
Near the interfaces, the initial turbulent length scale $\tilde{L}$ and intermittent factor $\gamma$ are specified as $0.055cm$ and $0$ for case A. For case B the initial $\tilde{L}$ is set to $0.015cm$, and $\gamma$ is set to $1$ due to its faster transition process than case A.   
\begin{figure}
    \centering
\includegraphics[width=0.55\textwidth]{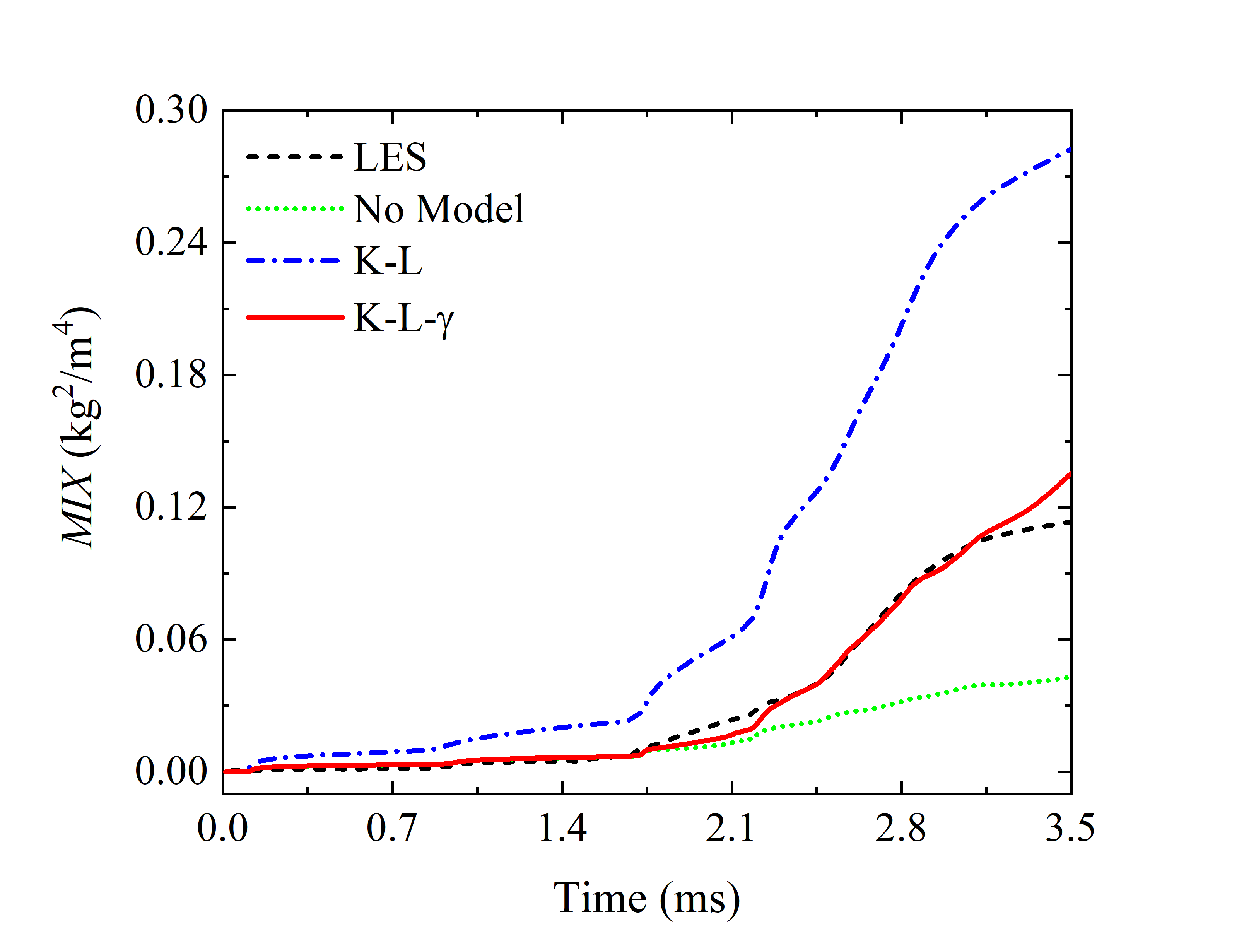}
    \caption{Total mixing (MIX) vs time for the inverse chevron case A.}
    \label{fig2}
\end{figure}
\begin{figure}
    \centering
\includegraphics[width=1\textwidth]{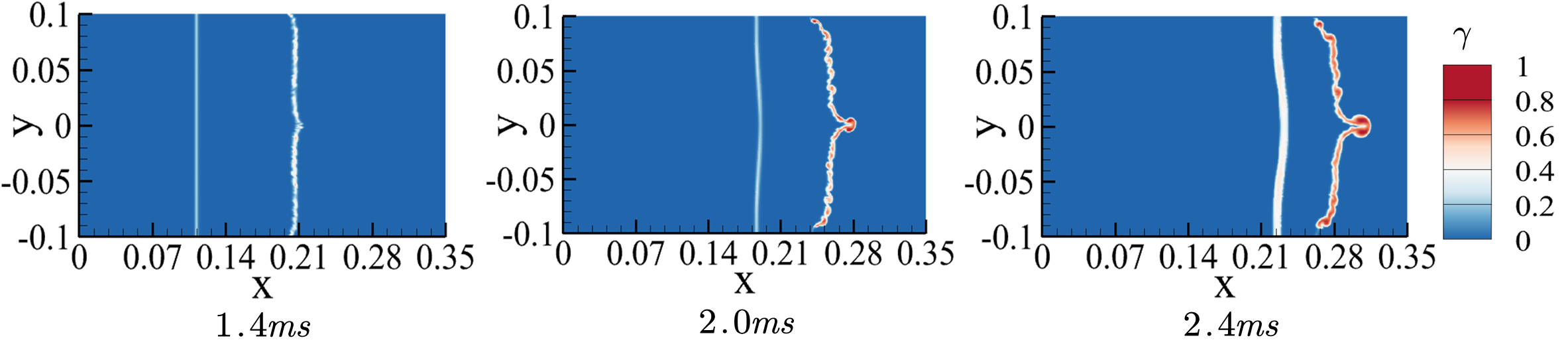}
    \caption{Contours of the $\gamma$ at three different moments. The subfigures from left to right represent the moments before the first reshocking ($1.4ms$), between the first and the second reshocking ($2.0ms$), and after the second reshocking ($2.4ms$), respectively.}\label{fig3}
\end{figure}

For mixing flows, the integral quantity $MIX$ is not prone to statistical noise and is a simple and effective way to measure variations of the total amount of mixing in complex flows \citep{Hahn2011Richtmyer}. The calculation of $MIX$ is given by:
\begin{equation} \label{eq19}
MIX=\int{\bar{\rho}}^2\tilde{Y}_{1}\tilde{Y}_{2}dV, 
\end{equation}
where $\tilde{Y}_{1}$ and $\tilde{Y}_{2}$ represent the mass fractions of the two species, with $\tilde{Y}_{1}\!=\!1-\tilde{Y}_{2}$.
For three-dimensional (3-D) simulations $dV\!=\!dxdydz$, whereas for 2-D simulations, $dV\!=\!dxdy$.

Fig.\ref{fig2} illustrates the temporal evolution of $MIX$ for the inverse chevron case. 
The 3-D LES is implemented using the explicit subgrid stress model recently proposed by \citet{Xiao2022exp}, which shows a good predictive accuracy for mixing problems induced by interfacial instabilities. 
To make the 2-D and 3-D simulations comparable, the results of LES are obtained by firstly averaging the 3-D flow field along the spanwise direction, and then calculating the $MIX$ based on (\ref{eq19}). 
More discussions about the results of the LES will be published separately.
In addition, two additional simulations are conducted to highlight the effect of the proposed model (K-L-$\gamma$), one of which not including any model (No Model) and the other using the baseline K-L model.

Fig.\ref{fig2} shows that, at approximately $0.1ms$ and $0.85ms$, the incident shock wave successively impacts on the  planar and chevron  interfaces,  resulting in a slight increase in $MIX$. 
At around $1.7ms$, a sudden growth in total mixing occurs when the mixing region is firstly reshocked by the reflected wave. Subsequently, around $2.2ms$, the mixing region is reshocked again, leading to a rapid rise in $MIX$.
The evolutionary process observed and the results of LES indicate that prior to reshocking, the mixing is at a low-mixed level. 
Consequently, the flow should be characterized as laminar or low-turbulent, which is accurately represented by the $\gamma$ contour  at $1.4ms$ in Fig.\ref{fig3}, where $\gamma$ exhibits consistently small values throughout the entire mixing region. Quantitatively, the K-L-$\gamma$ model also effectively predict the low-mixed state while the baseline model overpredicts the mixing evolution.              
After the mixing region is reshocked by the reflected wave, the flow undergoes a significant transition phase, during which the onset of transition is accurately identified by the K-L-$\gamma$ model.
Subsequently, the turbulence intensity enhances dramatically, and $\gamma$ promptly responds to these flow variations, exhibiting larger amplitudes within the mixing region, as shown in the contour at $2.0ms$ in Fig.\ref{fig3}.  
After the second reshocking event, the mixing enters a high-turbulent and well-mixed state. Such a flow variation is visually depicted by the contour of $\gamma$ at $2.4ms$ in Fig.\ref{fig3}, where $\gamma$ has a larger value and  the mixing region become  wider. The abovementioned mixing evolution is predicted accurately by the K-L-$\gamma$ model, demonstrating a good agreement with the LES results. Conversely, the K-L model  presents an absurd overprediction. 

\begin{figure}
\centering
\subfigure{
\includegraphics[width=0.47\textwidth,height=0.4\textwidth]{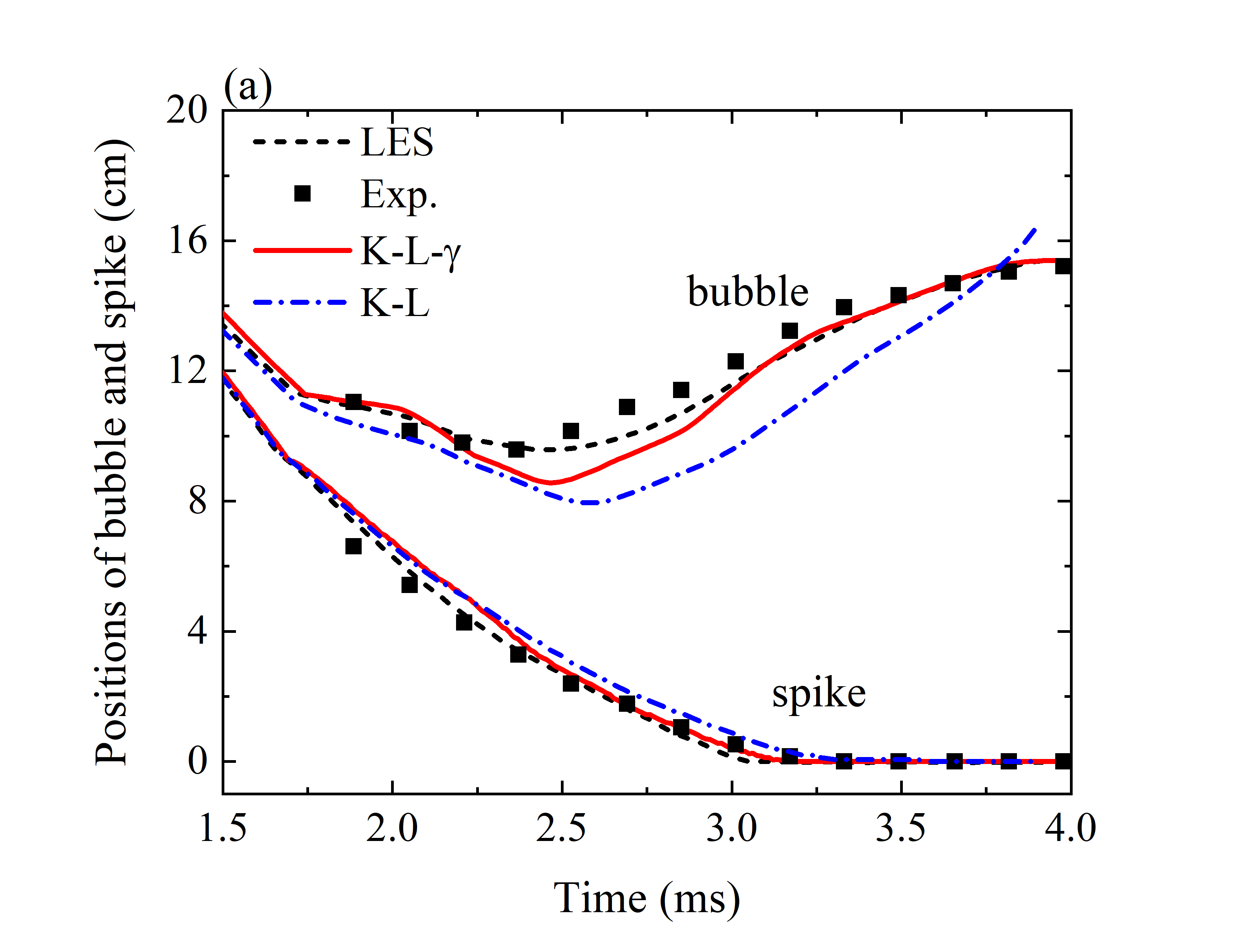}} \hspace{2mm}
\subfigure{
\includegraphics[width=0.47\textwidth,height=0.4\textwidth]{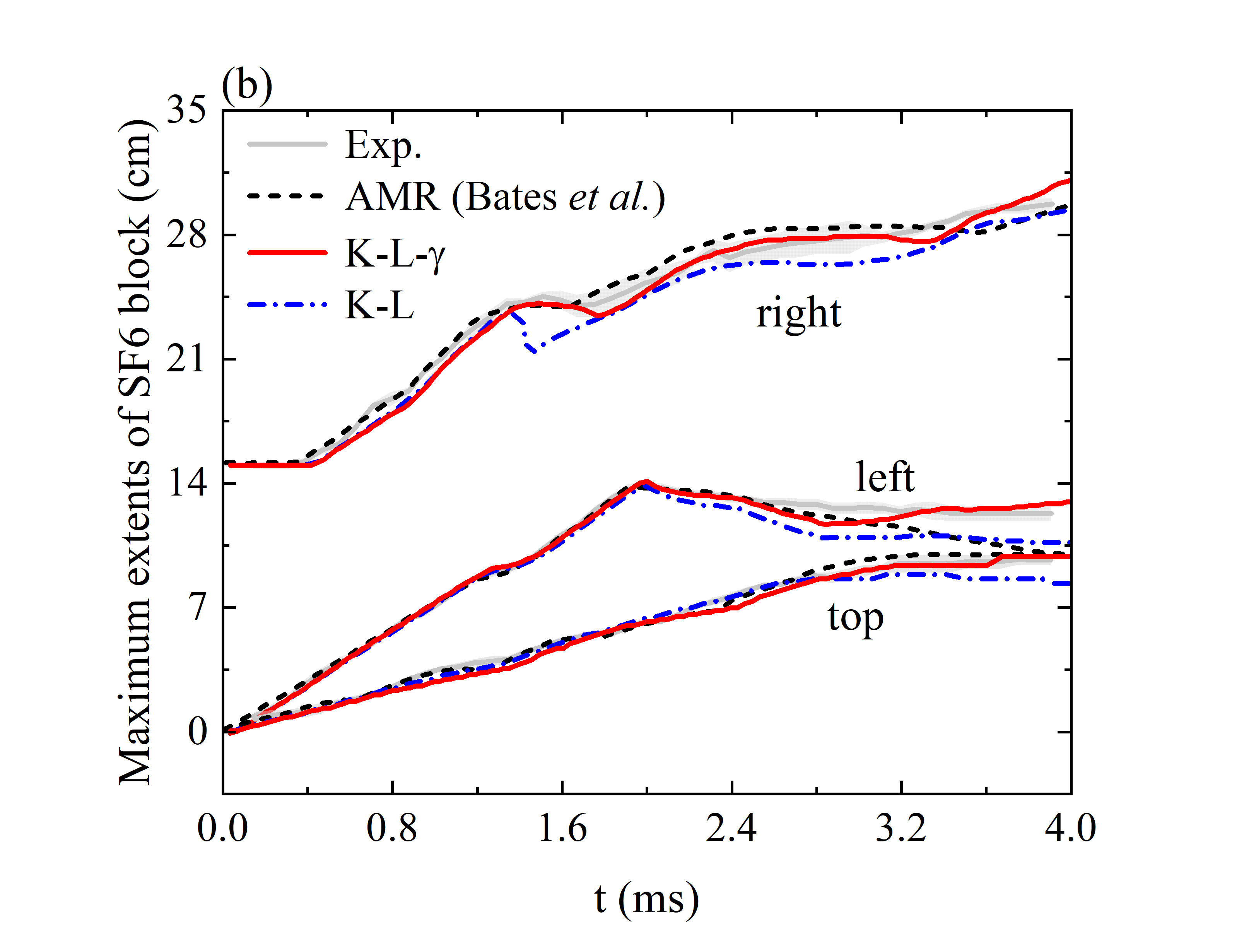}} 
\caption{(a) Distances of the characteristic spike and bubble from the end wall for case A. (b) Evolution of extends of the SF6 block along the x and y directions for case B.} \label{fig4}
\end{figure}


Fig.\ref{fig4}(a) further compares the evolution of distances between the end wall and the characteristic bubble and spike structures, which are tracked based on the mass fraction 0.5 at the down wall (left edge of the wall-bubble) and the SF6 mass fraction 0.1 (the tip of spike).
Results indicate that all the simulations accurately capture the location evolution of the spike's tip. Although the bubble position is difficult to predict, the proposed K-L-$\gamma$ model has a significantly better performance than the baseline model. 

The shock tube case shown in Fig.\ref{fig1}(b) is used to assess the generalization performance of the proposed model.
Fig.\ref{fig4}(b) shows the evolutions of the maximum extends of the SF6 block along the $x$ and $y$ directions. 
The experimental results are presented as error bands, 
while the Adaptive Mesh Refinement (AMR)  results are extracted from the high-resolution simulation conducted by \citet{Bates2007rich}.
The locations of the left, right and top sides of the SF6 block are determined based on the volume fraction of $0.1$. 
Fig.\ref{fig4}(b) indicates that the predictions of the K-L-$\gamma$ model largely fall within the error band provided by the experiment, demonstrating a significant improvement over the baseline K-L model.
Notably, around $1.6ms$, there is a noteworthy disparity between the predictions of the two RANS models with respect to the right position of the SF6 block. 
At this moment, the mixing region encounters a reshocking, which promotes the transition behaviour. 
The reshocking caused by the reflected wave compresses the mixing region, leading to a shift of the SF6 block towards the left side.
The experimental and AMR results indicate that the compression process is weak, as observed from the minimal variation in the right extent of the SF6 block. 
The present K-L-$\gamma$ model reasonably captures this compression behaviour and accurately predicts the evolution of the mixing transition stage.
In contrast, the K-L model performs poorly.

\section{Conclusions and discussions}\label{Sec:Conclusion and discussion}
The local spatio-temporal dependence of mixing transition requires that RANS models possess the corresponding capability to describe this characteristic.
Inspired by the intermittent factor widely used in boundary layer transition within the aerospace field, this study extends this well-developed modeling strategy to the mixing problems induced by interfacial instabilities.
Specifically, the concept of the intermittent factor $\gamma$ is defined based on the ensemble-averaged approach, and a transport equation is built for it. 
Subsequently, taking the well-developed K-L turbulent mixing model as an example, we couple the $\gamma$ equation into the K-L model to modify the two key source terms that dominate the mixing evolution, i.e. the Reynolds stress and buoyancy product terms. 
To validate the performance of the proposed model, two representative reshocked RM mixing cases are examined.
The good predictive results demonstrate that the present model can accurately capture the onset of the transition and depict its subsequent evolution. 
Furthermore, it exhibits the potential to provide accurate predictions for the entire process of mixing evolution.

To the best of our knowledge, it is the first study that an extra transport equation for intermittent factor has been proposed for a RANS mixing transition model.
More importantly, the current modeling framework exhibits flexibility and demonstrates a promising potential for more advanced modeling strategies of mixing transition. 
It is expected that the present framework can be extended to other turbulent mixing models, such as the K-$\epsilon$, BHR models, and so on. 

Nonetheless, several challenges remain to be addressed.
Firstly, additional cases are needed to extensively validate the performance of the proposed model. 
However, a scarcity of benchmark cases demonstrating notable transition effects exists, so further experiments and high-resolution simulations are desired to mitigate this limitation.
Secondly, the prediction for the $MIX$ in Fig.\ref{fig2} can be further improved, particularly during the initial stage of transition. This can be probably achieved by improving the model of Eq. (\ref{eq18}) to describe the growth rate of $\gamma$.

\backsection[Acknowledgements]
 
 This work was supported by the National Natural Science Foundation of China (Grant Nos. 12222203, 92152102, 11972093).

\backsection[Declaration of Interests]
 
 The authors report no conflict of interest.








\bibliographystyle{jfm}
\bibliography{jfm-template}






\end{document}